\documentclass[preprint,aps,showpacs,showkeys]{revtex4}

\usepackage{epsfig,amsmath,amssymb,color}
\begin{document}

\title{Quantum Heisenberg antiferromagnet on low-dimensional frustrated lattices}

\author{Oleg Derzhko}
\affiliation{Institute for Condensed Matter Physics,
             National Academy of Sciences of Ukraine,
             1 Svientsitskii Street, L'viv-11, 79011, Ukraine}
\affiliation{Department for Theoretical Physics,
             Ivan Franko National University of L'viv,
             12 Drahomanov Street, L'viv-5, 79005, Ukraine}

\author{Taras Krokhmalskii}
\affiliation{Institute for Condensed Matter Physics,
             National Academy of Sciences of Ukraine,
             1 Svientsitskii Street, L'viv-11, 79011, Ukraine}
\affiliation{Department for Theoretical Physics,
             Ivan Franko National University of L'viv,
             12 Drahomanov Street, L'viv-5, 79005, Ukraine}

\author{Johannes Richter}
\affiliation{Institut f\"ur Theoretische Physik,
             Universit\"at Magdeburg,
             P.O. Box 4120, 39016 Magdeburg, Germany}

\date{March 20, 2011}

\begin{abstract}
Using a lattice-gas description of the low-energy degrees of freedom 
of the quantum Heisenberg antiferromagnet on the frustrated two-leg ladder and bilayer lattices
we examine the magnetization process at low temperatures for these spin models.
In both cases the emergent discrete degrees of freedom implicate a close relation 
of the frustrated quantum Heisenberg antiferromagnet 
to the classical lattice gas with finite nearest-neighbor repulsion 
or, equivalently, to the Ising antiferromagnet in a uniform magnetic field. 
Using this relation we obtain analytical results for thermodynamically large systems in the one-dimensional case.
In the two-dimensional case we perform classical Monte Carlo simulations for systems of up to $100 \times 100$ sites.
\end{abstract}



\maketitle

%\renewcommand\baselinestretch{1.0}
%\large\normalsize

\section{Introductory remarks}
\label{sec1}
\setcounter{equation}{0}

Experimental and theoretical studies of magnetization processes is a hot topic of the modern condensed matter physics.
Although a common wisdom says that the magnetization $M$ monotonically increases approaching its saturation value
as the applied field $h$ increases,
experimental observations for many materials often demonstrate a nontrivial dependence of $M$ on $h$ at low temperatures,
which may show plateaus and jumps 
(see, e.g., Ref. \cite{honecker}).
This kind of behavior may be caused by competing interactions which occur in frustrated quantum systems.
The theoretical description of the magnetization processes in frustrated quantum spin systems 
has attracted a lot of interest during the last decades,
and theoretical predictions for the low-temperature magnetization curves of thermodynamically large systems
are of great interest \cite{honecker}.

On the other hand,
recently the concept of independent localized-magnon states 
for the quantum Heisenberg antiferromagnet on certain classes of frustrated lattices
has been introduced \cite{lm1}.
Later on it has been successfully used to describe magnetothermodynamics of these systems \cite{lm2}.
The independent localized-magnon states are the ground states in strong magnetic fields
and therefore the independent localized-magnon picture 
(or hard-core model description)
is adequate at the strong-field low-temperature regime.
Moreover, very recently, the independent localized-magnon description has been improved:
For some lattices \cite{independent_lm} it appears possible to take into account the low-energy excited states 
(interacting localized-magnon states) too \cite{interacting_lm}.
As a result, 
a lattice-gas description of low-energy degrees of freedom of certain frustrated quantum spin systems has been elaborated.
In Ref. \cite{interacting_lm} some results for thermodynamic quantities 
(mainly concerning temperature dependences of specific heat, staggered susceptibility and entropy) 
have been reported too.
The aim of the present paper is to use the lattice-gas description to study magnetization processes at low temperatures 
which have not been considered so far.
We shall demonstrate that such a description for the considered frustrated quantum magnets
yields very accurate results for the low-temperature quantities.

The paper is organized as follows.
In the next section (Sec.~\ref{sec2}) we define the quantum spin models and explain how the lattice-gas picture emerges.
Then, in Sec.~\ref{sec3}, we calculate the magnetization of the frustrated quantum spin systems 
considering separately the one-dimensional case and the two-dimensional case.
We report results for thermodynamically large systems 
obtained using transfer-matrix method (one-dimensional case) and classical Monte Carlo simulations (two-dimensional case).
We discuss in detail the low-temperature uniform magnetization and susceptibility 
within the frames of the lattice-gas description.
Finally, in Sec.~\ref{sec4}, we summarize our findings.

\section{Frustrated quantum spin systems. Lattice-gas description}
\label{sec2}
\setcounter{equation}{0}

\begin{figure}
\begin{center}
\includegraphics[clip=on,width=5.75cm]{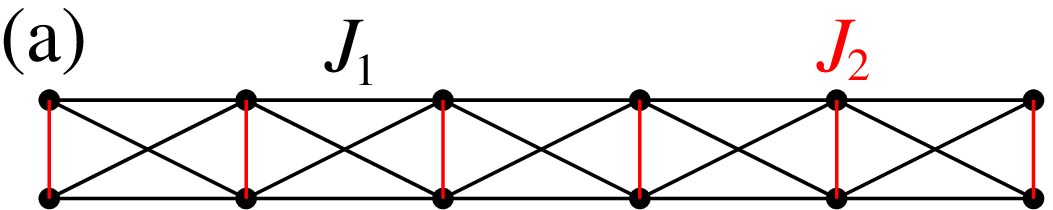}
\includegraphics[clip=on,width=7.75cm]{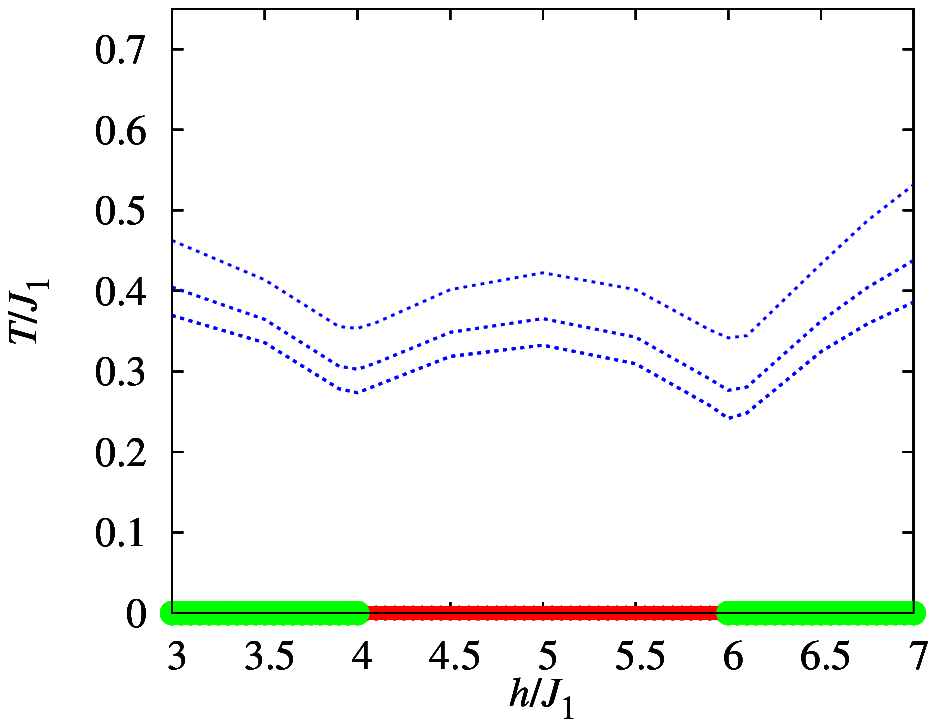}
\\
\includegraphics[clip=on,width=5.75cm]{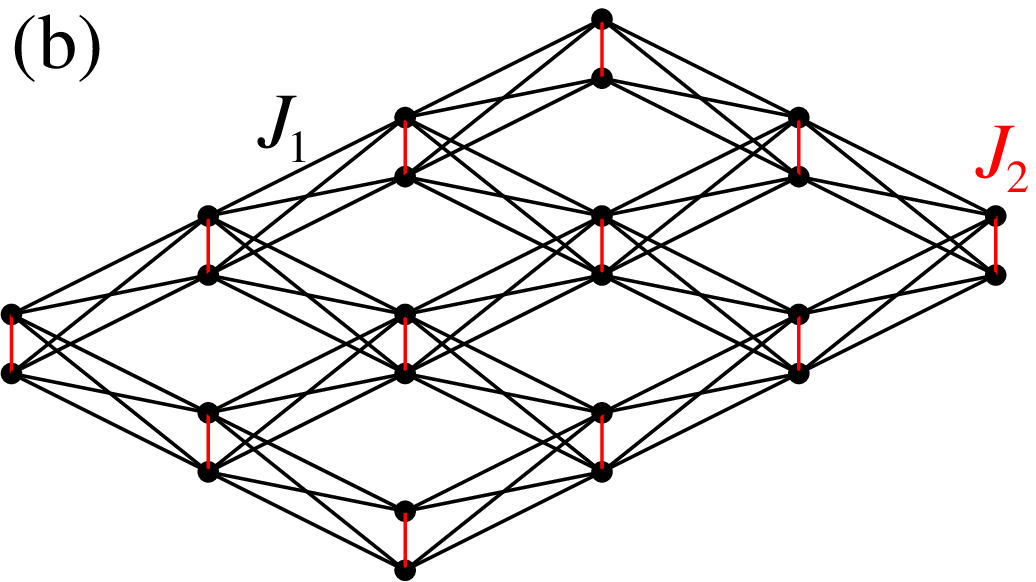}
\includegraphics[clip=on,width=7.75cm]{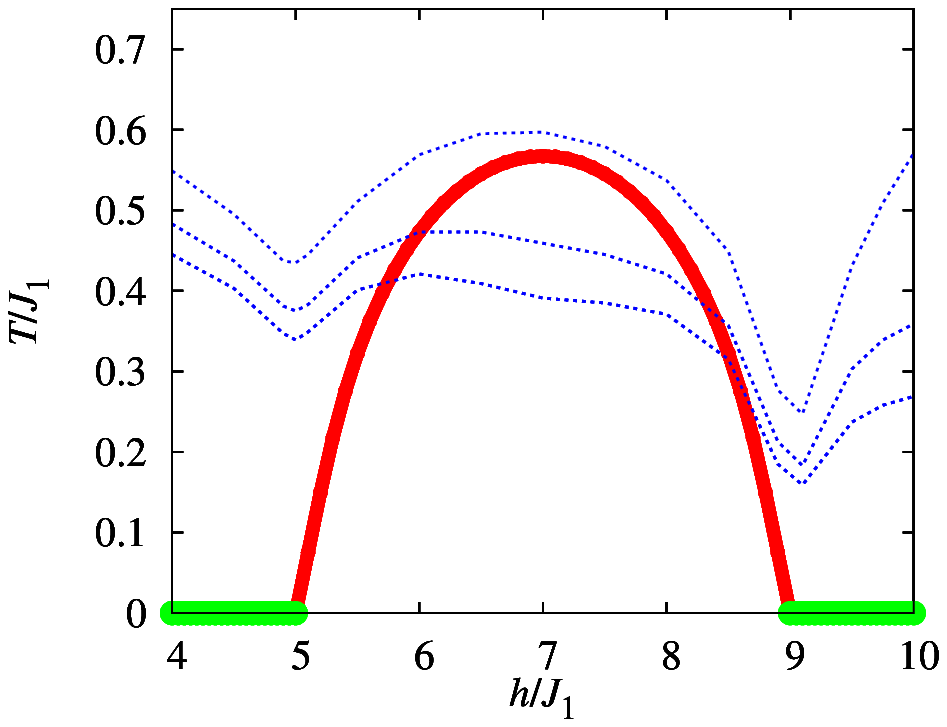}
\end{center}
\caption{The two lattices considered in the paper: the frustrated two-leg ladder (a) and the frustrated bilayer (b).
We also show the phase diagrams as they follow from the lattice-gas description
of the $s=1/2$ Heisenberg antiferromagnet in a magnetic field 
on the frustrated two-leg ladder lattice with $J_1=1$, $J_2=4$  ($h_2=4$, $h_1=6$)
and 
on the frustrated bilayer lattice with $J_1=1$, $J_2=5$  ($h_2=5$, $h_1=9$).
A staggered occupation of vertical bonds by localized magnons 
(a kind of ``antiferromagnetic'' long-range order) 
occurs along the red line between $h_2$ and $h_1$ at $T=0$ (ladder)
or 
below the red critical line $T_c(h)$ with starting and end points at $h_2$, $T=0$ and $h_1$, $T=0$ (bilayer), 
whereas a uniform occupation of vertical bonds 
(a kind of ``ferromagnetic'' long-range order) 
occurs along the green lines $h<h_2$ at $T=0$ and $h_1<h$ at $T=0$.
The remaining part of the phase diagrams corresponds to a disordered phase.
Moreover,
we show the lines below which the exact diagonalization data and the lattice-gas predictions 
for the specific heat of the finite system of $N=16$ sites coincide with the accuracy up to 5\%, 2\%, and 1\% 
(blue dashed lines from top to bottom).}
\label{f1}
\end{figure}

We consider $N=2{\cal{N}}$ quantum spins 1/2 placed on the two lattices shown in Fig.~\ref{f1},
see also Refs. \cite{ho_mi_tr,bilayer}.
The Hamiltonian of the model reads 
\begin{eqnarray}
\label{2.01}
H=\sum_{(pq)}J_{pq}{\bf{s}}_p\cdot {\bf{s}}_q - hS^z,
\end{eqnarray}
where the sum runs over the bonds connecting neighboring sites on the lattice,
$J_{pq}>0$ is the antiferromagnetic interaction between the sites $p$ and $q$,
$J_{pq}$ takes two values, $J_2$ for vertical bonds and $J_1$ for horizontal and diagonal bonds,
$h\ge 0$ is the external magnetic field,
$S^z=\sum_{p=1}^Ns_p^z$ is the $z$ component of the total spin of the system.
In our study we impose periodic boundary conditions and usually set $J_1=1$ to fix the units.

It is useful to introduce the total spin on a vertical bond  
${\bf{t}}_m={\bf{s}}_{m,1}+{\bf{s}}_{m,2}$.
Here $m=1,\ldots,{\cal{N}}$ enumerates vertical bonds and runs over lattice sites of 
a simple chain in the case of the frustrated two-leg ladder 
or 
a square lattice in the case of the frustrated bilayer.
Then the Hamiltonian (\ref{2.01}) becomes
\begin{eqnarray}
\label{2.02}
H=\sum_{m}\left[\frac{J_2}{2}\left({\bf{t}}^2_m-\frac{3}{2}\right)-ht_m^z\right]+J_1\sum_{(ml)}{\bf{t}}_m\cdot {\bf{t}}_l,
\end{eqnarray}
where the second sum runs over the nearest-neighbor bonds on the simple chain or the square lattice.
It is evident from Eq. (\ref{2.02}), 
that the total spin of each vertical bond $t_m$, $m=1,\ldots,{\cal{N}}$, ${\bf{t}}_m^2=t_m(t_m+1)$, 
is a good quantum number. 
Hence, the Hamiltonian (\ref{2.02}) depends on the set of quantum numbers $\{ t_m\}$, $t_m=0,1$.
As a result, the properties of the model can be studied in much more detail.
In particular, 
a subset of $2^{\cal{N}}={1.41\ldots}^N$ low-lying eigenstates of the total set of $2^N$ eigenstates 
can be constructed exactly \cite{interacting_lm,ho_mi_tr}.
Moreover, 
their contribution to thermodynamics can be calculated using a classical lattice-gas model \cite{interacting_lm}.

We start with an illustration  of the eigenstates which we will consider in more detail. 
First we consider states which consist of $n$ singlets on the vertical bonds $m_1,\ldots,m_n$ (i.e., ${\bf{t}}_{m_i}^2=0$)  
and  ${\cal N}-n$ fully polarized triplets (${\bf{t}}_m^2=2$, $t_m^z=1$) on the remaining  vertical bonds. 
We impose a hard-core rule,
i.e., the occupation of neighboring vertical bonds by singlets is forbidden.
Following the notations introduced in Refs. \cite{lm1,lm2,independent_lm} 
we call these states ``independent localized-magnon states'' 
and a singlet on a vertical bond a ``localized magnon''.
The energy of these independent localized-magnon states is $E_{n}^{\rm{lm}}=E_{\rm{FM}}-n\epsilon_1$,
where 
$E_{\rm{FM}}={\cal{N}}J_1+{\cal{N}}J_2/4$,
$\epsilon_1=J_2+2J_1=h_1$
(one-dimensional case) and 
$E_{\rm{FM}}=2{\cal{N}}J_1+{\cal{N}}J_2/4$,
$\epsilon_1=J_2+4J_1=h_1$
(two-dimensional case).
The degeneracy of the independent localized-magnon states $g_{\cal{N}}(n)$ 
equals to the canonical partition function ${\cal{Z}}_{\rm{hc}}(n,{\cal{N}})$ 
of $n$ hard-core objects (hard dimers or hard squares) on the lattice (simple chain or square lattice) of $\cal{N}$ sites.
It is important to mention 
that the independent localized-magnon states are the ground states in the subspaces with $S^z={\cal{N}}-1,\ldots,{\cal{N}}/2$ 
if $J_2/J_1\ge 2$ (one-dimensional case) or $J_2/J_1\ge 4$ (two-dimensional case), cf. Ref. \cite{HJS},
and that these states are linearly independent \cite{linear}.

Next we relax the hard-core rule, 
i.e., we allow the localized magnons (singlets on vertical bonds) to be nearest neighbors.
According to Ref. \cite{interacting_lm} we call these states ``interacting localized-magnon states''.
Any  pair of neighboring localized magnons increases the energy by $J_1$. 
As a result, 
for a given $S^z={\cal{N}}-n$, $n=1,\ldots,{\cal{N}}/2$ 
the energies of these interacting localized-magnon excited states become $E_n^{\nu}=E_{n}^{\rm{lm}}+\nu J_1$, 
where $\nu$ is the number of pairs of neighboring localized magnons, 
cf. also Ref. \cite{interacting_lm}.
The interacting localized-magnon states are the low-lying excited states
for  $S^z={\cal{N}}-n$, $n=1,\ldots,{\cal{N}}/2$ in the strong-coupling regime, 
i.e., when $J_2/J_1$ is sufficiently large.
More precisely,
from exact diagonalization data for finite systems we have found that the strong-coupling regime occurs 
when $J_2>J_2^c$ with
$J_2^c/J_1\approx 3.00$ (one-dimensional case)
and
$J_2^c/J_1\approx 4.65$ (two-dimensional case) \cite{interacting_lm}.

For lower values of the magnetization, $S^z={\cal{N}}/2-r$, $r=1,\ldots,{\cal{N}}/2$, 
where no independent localized-magnon states exist,
the class of interacting localized-magnon states contains the ground-state manifold as well as low-lying excited states 
in the strong-coupling regime.
The ground-state manifold is built by $n={\cal{N}}/2+r$, $r=1,\ldots,{\cal{N}}/2$ localized magnons, 
where, e.g., ${\cal{N}}/2$ magnons occupy one sublattice of the underlying lattice (simple chain or the square lattice) completely,
and the remaining $r$ localized magnons sit on the other sublattice.
The energy of this state is $E_{{\cal{N}}/2+r}=-{\cal{N}}J_2/4-rJ_2$, $J_2=h_2$.
The low-lying excited states are constructed from the ground state 
by rearranging the localized magnons to increase the number of neighboring magnons. 
Then each new pair of neighboring localized magnons increases the energy by $J_1$.

Note that the interacting localized-magnon states can be visualized as partially overlapping hard-core objects
(in contrast to the independent localized-magnon states which can be visualized as nonoverlapping hard-core objects).
Although the degeneracy of the interacting localized-magnon states can be also calculated 
in terms of a canonical partition function of a system of hard-core objects,
the required contribution of the independent and interacting localized-magnon states 
to the partition function $Z(T,h,N)$ of the quantum spin system (\ref{2.01}) is conveniently taken into account 
within the frames of a lattice-gas model of classical particles with finite nearest-neighbor repulsion $V=J_1$,
see Ref. \cite{interacting_lm}.

In the strong-coupling regime, 
when the constructed independent and interacting localized-magnon states 
dominate the partition function $Z(T,h,N)$ of the quantum spin system (\ref{2.01}) at low temperatures,
for $Z(T,h,N)$ we can write \cite{interacting_lm}
\begin{eqnarray}
\label{2.03}
Z(T,h,N)
\approx Z_{\rm{LM}}(T,h,N)
=\sum_{n_1=0,1}\ldots\sum_{n_{\cal{N}}=0,1}
e^{-\frac{E_{\rm{FM}}-h{\cal{N}}+(h-h_1)\sum_{m}n_m+J_1\sum_{(ml)}n_mn_l}{T}}
\nonumber\\
=e^{-\frac{E_{\rm{FM}}-h{\cal{N}}}{T}} \Xi_{\rm{lg}}(T,\mu,{\cal{N}}),
\end{eqnarray}
where 
$\Xi_{\rm{lg}}(T,\mu,{\cal{N}})=\sum_{n_1=0,1}\ldots\sum_{n_{\cal{N}}=0,1}e^{-{\cal{H}}(\{n_m\})/T}$ 
is the grand-canonical partition function of the lattice gas with the Hamiltonian
${\cal{H}}(\{n_m\})=-\mu\sum_m n_m+J_1\sum_{(ml)}n_mn_l$, $\mu=h_1-h$.
Introducing the on-site spin variables $\sigma_m=\pm 1$ according to the relations 
$\sigma_m=2n_m-1$ and $n_m=(1+\sigma_m)/2$ 
one arrives at the Hamiltonian of the antiferromagnetic Ising model in a uniform magnetic field.

In Ref. \cite{interacting_lm} we have performed extensive studies of the relation 
between the frustrated quantum spin model (\ref{2.01}) 
and the classical lattice-gas model with the Hamiltonian ${\cal{H}}(\{n_m\})$
for finite systems (up to $N=32$)
to clarify to what extent the latter model can reproduce the thermodynamic properties of the former one
in the strong-coupling regime.
In particular, 
we have checked the energies and degeneracies of low-lying states for $S^z=N/2,\ldots,0$,
have compared the field and temperature dependences of various thermodynamic quantities,
have estimated the temperature range 
until which the lattice-gas predictions are in an excellent agreement with exact diagonalization data.
Thus, 
for the frustrated two-leg ladder with $J_1=1$, $J_2=4$ 
(characteristic fields are $h_2=4$ and $h_1=6$)
and 
for the frustrated bilayer with $J_1=1$, $J_2=5$
(characteristic fields are $h_2=5$ and $h_1=9$)
we have found that the lattice-gas picture works perfectly well at least up to $T=0.5$
(and even up to $T=1$ if $J_2$ has larger value $J_2=10$).
To supplement this conclusion of Ref. \cite{interacting_lm}
we perform a new comparison of the exact diagonalization data and the lattice-gas predictions 
for the specific heat of the finite system of $N=16$ sites.
Our results are shown in the phase diagrams in Fig.~\ref{f1}.
We find that below the dashed lines in the phase diagrams in Fig.~\ref{f1} both results coincide 
with the accuracy up to 5\%, 2\%, and 1\% (from top to bottom).

The considered frustrated quantum spin systems exhibit an interesting low-temperature behavior 
if the magnetic field $h$ is between $h_2$ and $h_1$
which is related to an ordering of localized magnons on the simple chain or the square lattice.
Both these lattices are bipartite ones,
i.e., 
they consist of two sublattices $A$ and $B$ 
and any neighboring sites always belong to the sublattice $A$ and the sublattice $B$.
In the two-dimensional case the filling of the sublattices by localized magnons realizes as an order-disorder phase transition,
which has a direct analogy to the phase transition for the square-lattice Ising antiferromagnet in a uniform magnetic field
and belongs to the two-dimensional Ising model universality class.
There are many studies on the square-lattice Ising antiferromagnet in a uniform magnetic field, 
see, e.g., Refs. \cite{mueller-hartman,wu,wang,betts},
and we may borrow the existing knowledge 
to examine the two-dimensional frustrated bilayer quantum Heisenberg antiferromagnet.
In particular, 
a phase diagram in the half-plane ``magnetic field -- temperature'' 
of the square-lattice Ising antiferromagnet in a uniform magnetic field 
has been discussed
and
a critical line $T_c(h)$ which separates the antiferromagnetically ordered phase and the disordered phase 
has been calculated using different approximations.
This phase diagram in the context of the frustrated bilayer has been reproduced 
in our classical Monte Carlo simulations \cite{interacting_lm}, see Fig.~\ref{f1}b.
While crossing the critical line, 
the thermodynamic quantities for the frustrated bilayer exhibit singularities:
The specific heat shows a logarithmic singularity,
the staggered magnetization 
(which may play a role of the order parameter) 
decays within the (``antiferromagnetically'') ordered phase with the exponent $\beta=1/8$,
the staggered susceptibility diverges with the exponent $\gamma=7/4$.

Using the elaborated lattice-gas picture we will calculate in the present paper 
the uniform magnetization $M(T,h,N)$ and  the uniform susceptibility $\chi(T,h,N)=\partial M(T,h,N)/\partial h$, 
which are both important and easily accessible quantities for experimental studies.

\section{Uniform magnetization and susceptibility}
\label{sec3}
\setcounter{equation}{0}

We use Eq. (\ref{2.03}) to calculate $M(T,h,N)$ and $\chi(T,h,N)$ for $h\ge 0$ at low temperatures $T$  
for the frustrated quantum Heisenberg antiferromagnet (\ref{2.01}) on the considered lattices. 
Using standard formulas for $M$ and $\chi$ we get
\begin{eqnarray}
\label{3.01}
M(T,h,N)
=-\frac{\partial}{\partial h}\left[-T\ln Z(T,h,N)\right]
={\cal{N}}-\overline{n},
\;\;\;
\overline{n}=T\frac{\partial\ln\Xi_{\rm{lg}}(T,\mu,{\cal{N}})}{\partial\mu},
\nonumber\\
\chi(T,h,N)
=\frac{\partial M(T,h,N)}{\partial h}
=\frac{\partial \overline{n}}{\partial \mu}
=\frac{1}{T}\left(\overline{n^2}-{\overline{n}}^2\right),
\end{eqnarray}
where 
$\overline{(\ldots)}=\sum_{n_1=0,1}\ldots\sum_{n_{{\cal{N}}}=0,1}(\ldots)e^{-{\cal{H}}(\{n_m\})/T}/\Xi_{\rm{lg}}(T,\mu,{\cal{N}})$
denotes the grand-canonical average for the classical lattice-gas model
and 
$n=\sum_{m=1}^{\cal{N}}n_m$.

\subsection{Frustrated two-leg ladder}

In the one-dimensional case we use the transfer-matrix approach \cite{baxter} 
to obtain explicitly $\Xi_{\rm{lg}}(T,\mu,{\cal{N}})$
and hence $\overline{n}$ and $\partial\overline{n}/\partial\mu$,
see Eq. (\ref{3.01}).
After somewhat lengthy but straightforward calculations
we obtain the following final results in the thermodynamic limit ${\cal{N}}\to\infty$
\begin{eqnarray}
\label{3.02}
\frac{M(T,h,N)}{{\cal{N}}}=1-\frac{c_+}{\lambda_+},
\nonumber\\
\frac{T\chi(T,h,N)}{{\cal{N}}}
=\frac{d_+}{\lambda_+}-\frac{c_+^2}{\lambda_+^2},
\nonumber\\
\lambda_+=\frac{1+w+\sqrt{(1-w)^2+4z}}{2},
\nonumber\\
c_+=\frac{2z-w(1-w)}{2\sqrt{(1-w)^2+4z}}+\frac{w}{2},
\nonumber\\
d_+=\frac{2z-w(1-2w)}{2\sqrt{(1-w)^2+4z}}-\frac{[2z-w(1-w)]^2}{2[(1-w)^2+4z]^{\frac{3}{2}}}+\frac{w}{2}
\end{eqnarray}
with $z=e^{(h_1-h)/T}$, $h_1=J_2+2J_1$, $w=ze^{-J_1/T}$.

\begin{figure}
\begin{center}
\includegraphics[clip=on,width=8cm]{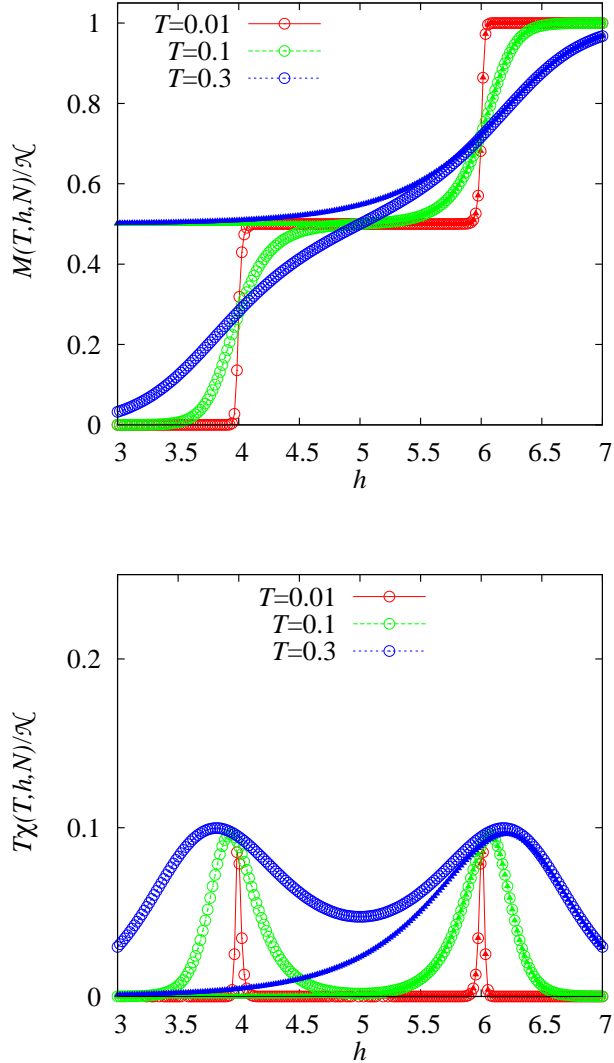}
\end{center}
\caption{Transfer-matrix approach results for the frustrated two-leg ladder (\ref{2.01}) 
with $J_1=1$, $J_2=4$, ${\cal{N}}\to\infty$: 
$M/{\cal{N}}$ vs $h$ (upper panel) and $T\chi/{\cal{N}}$ vs $h$ (lower panel) 
at different temperatures $T=0.01,\,0.1,\,0.3$.
Small filled triangles correspond to the hard-dimer-model predictions valid around $h_1$ at very low temperatures only.}
\label{f2}
\end{figure}

Some results for $M$ vs $h$ and $\chi$ vs $h$ at low temperatures based on Eq. (\ref{3.02}) 
are collected in Fig.~\ref{f2} and discussed in Sec.~\ref{discu}.

\subsection{Frustrated bilayer}

The two-dimensional case is more complicated 
since we do not have an analytical solution for the square-lattice lattice-gas model with nearest-neighbor repulsion.
Therefore we use the formulas expressing $M(T,h,N)$ and $\chi(T,h,N)$ 
in terms of the grand-canonical averages $\overline{n}$ and $\overline{n^2}$,
see Eq. (\ref{3.01}),
and compute $\overline{n}$ and $\overline{n^2}$ using classical Monte Carlo simulations for large systems.
Namely,
we consider systems of ${\cal{N}}=50 \times 50$ and $100 \times 100$ sites,
exploit the usual Metropolis algorithm,
and perform 220 000 Monte Carlo steps.

\begin{figure}
\begin{center}
\includegraphics[clip=on,width=8cm]{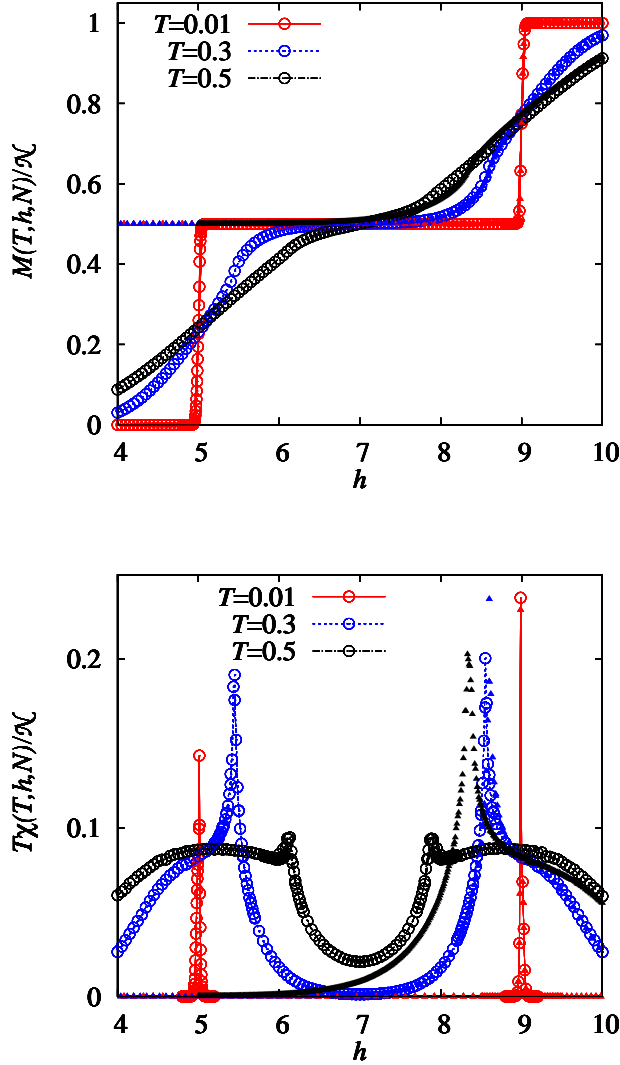}
\end{center}
\caption{Classical Monte Carlo simulation results for the frustrated bilayer (\ref{2.01}) 
with $J_1=1$, $J_2=5$, ${\cal{N}}=50\times 50$ and $100 \times 100$: 
$M/{\cal{N}}$ vs $h$ (upper panel) and $T\chi/{\cal{N}}$ vs $h$ (lower panel) 
at different temperatures $T=0.01,\,0.3,\,0.5$.
Small filled triangles correspond to the hard-square-model predictions valid around $h_1$ at very low temperatures only.}
\label{f3}
\end{figure}

\begin{figure}
\begin{center}
\includegraphics[clip=on,width=8cm]{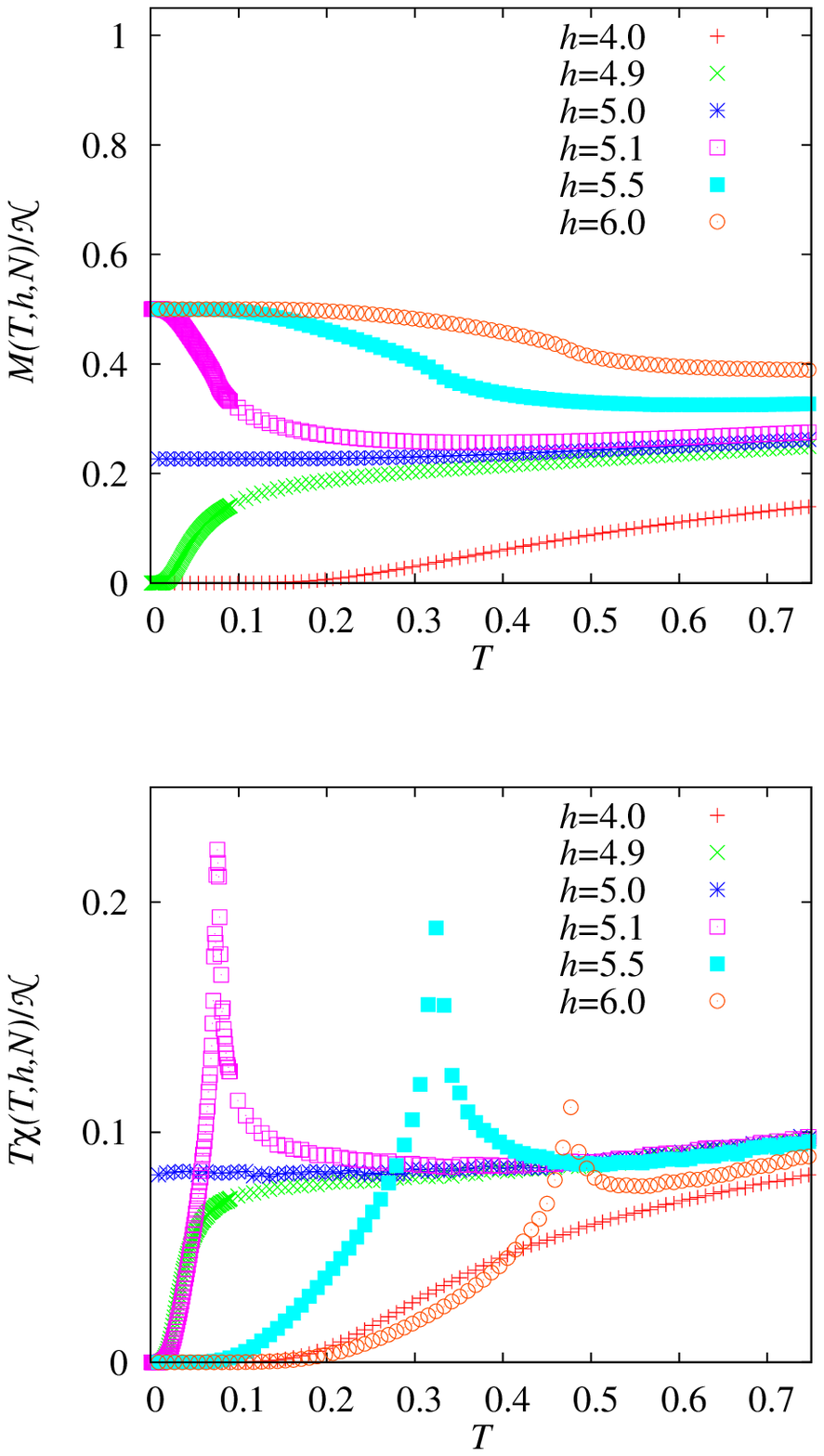}
\includegraphics[clip=on,width=8cm]{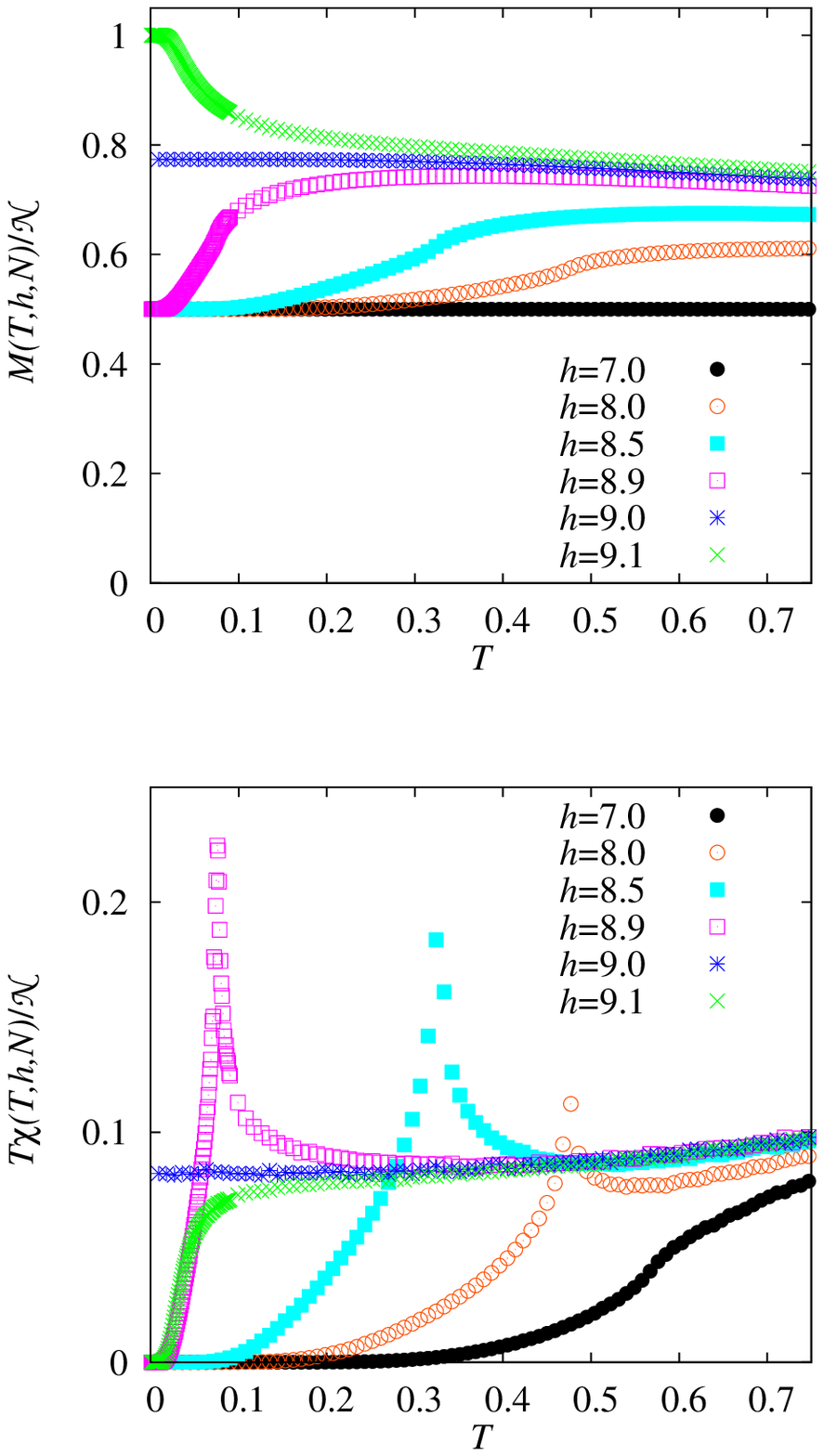}
\end{center}
\caption{Classical Monte Carlo simulation results for the frustrated bilayer (\ref{2.01}) 
with $J_1=1$, $J_2=5$, ${\cal{N}}=50 \times 50$ and $100 \times 100$: 
$M/{\cal{N}}$ vs $T$ (upper panels) and $T\chi/{\cal{N}}$ vs $T$ (lower panels)
at different fields $h=4.0,\ldots,9.1$.}
\label{f4}
\end{figure}

Some results for $M$ vs $h$ and $\chi$ vs $h$ at low temperatures 
based on the Monte Carlo simulations of $\overline{n}$ and $\overline{n^2}$ for the lattice-gas model
are collected in Fig.~\ref{f3}.
We also present temperature dependences of $M$ and $\chi$ at different fields $h$ 
as they follow from the Monte Carlo simulations for the lattice-gas model,
see Fig.~\ref{f4}.
The reported results are discussed in Sec.~\ref{discu}.

\subsection{Discussion}
\label{discu}

Let us discuss the obtained results for the low-temperature magnetization processes 
for the two considered frustrated quantum Heisenberg antiferromagnets.
At zero temperature $T=0$ the magnetization curve $M(h)$ is very simple, 
namely, it consists only of plateaus and jumps, 
i.e.,
$M(T=0,h,N)/{\cal{N}}=1$, 1/2 or 0
if
$h_1<h$, $h_2<h<h_1$ or $h<h_2$, cf. Ref. \cite{ho_mi_tr}.
The resulting susceptibility $\chi(T=0,h,N)$ is zero in the plateau states, 
i.e., for all values $h \ne h_2$ and $h \ne h_1$. 
The jumps in $M(T=0,h,N)$ at $h=h_2$ and $h=h_1$ lead to $\delta$-singularities in the susceptibility.
This behavior of the zero-temperature magnetization curve 
reflects the change of the ground states with varying $h$,
i.e., 
completely empty two sublattices for $h_1<h$, 
one completely occupied with localized magnons sublattice and one completely empty sublattice for 
$h_2<h<h_1$ 
or 
completely occupied both sublattices with localized magnons for $h<h_2$.
At $h=h_1$ 
the energy of the (ground) states with different number of (independent) localized magnons is the same
resulting in a jump in the magnetization curve.
At $h=h_2$ 
the ground-state energy of the states in the subspaces with $S^z={\cal{N}}/2-r$, $r=1,\ldots,{\cal{N}}/2$ is also the same
that provides another jump in the magnetization curve and a $\delta$-singularity in the dependence $\chi(T=0,h,N)$ vs $h$ at $h=h_2$.

At nonzero temperatures $T>0$ the lattice-gas picture implies different scenario for the one-dimensional and two-dimensional cases.
For the frustrated two-leg ladder the occupation of both sublattices with localized magnons increases as $h$ decreases
(leading to a decrease of $M$),
however, for finite temperatures the two sublattices are equally occupied with localized magnons.
A smooth decrease of $M$ with two steep parts around $h_1$ and $h_2$ 
leads to two peaks in the susceptibility $\chi$ at low temperatures,
see Fig.~\ref{f2}.

The situation for the frustrated bilayer is different 
due to the existence of a long-range ordered phase at $T/J_1<T_0/J_1=1/[2\ln(\sqrt{2}+1)]\approx 0.567\,296 $ 
(see the phase diagram in Fig.~\ref{f1}b).
For a given temperature $T<T_0$ by variation of the magnetic field 
(that corresponds to a horizontal line in the phase diagram in Fig.~\ref{f1}b) 
the system undergoes two phase transitions at $h_1(T)$ and $h_2(T)$, where the line $T=const.$ crosses the critical line $T_c(h)$.
[Clearly, $h_1(T=0)=h_1$ and $h_2(T=0)=h_2$.]
For the fields between $h_1(T)$ and $h_2(T)$ there is a difference in the occupation 
of the two sublattices, 
i.e., we have a kind of ``antiferromagnetic'' long-range order of the localized magnons.
The total density of the localized magnons related to the uniform magnetization $M$ exhibits a kind of inflection,
see the upper panel in Fig.~\ref{f3}, 
which results in a sharp peak in $\chi$,
see the lower panel in Fig.~\ref{f3}. 
An infinite slope in $M(h)$ at $h_1(T)$ and $h_2(T)$ resulting in the $\delta$-peaks in $\chi(h)$ occurs at $T=0$,
whereas at finite temperatures $T$ the well-pronounced peaks are finite, i.e., there is no divergency of $\chi$.
An alternative way to cross the critical line is to fix $h$ at $h_2< h <h_1$ and to vary the temperature $T\ge 0$.
Corresponding profiles for $M$ and $T\chi$ are given in Fig.~\ref{f4}.
Recall that the zero-temperature value of magnetization 
$M(T=0,h,N)/{\cal{N}}$ is 1 if $h_1<h$, 1/2 if $h_2<h<h_1$, and 0 if $h<h_2$.
The susceptibility $\chi$ is strongly suppressed in the ``antiferromagnetically'' ordered phase. 
That indicates that the zero-temperature plateau state between $h_2$ and $h_1$ extends to finite temperatures 
(note that this statement does not hold for the plateau state at $h<h_2$ and $h_1<h$).
Again the temperature profiles of $\chi$ indicate the critical line $T_c(h)$, 
see the lower panels in Fig.~\ref{f4}. 

Let us finally briefly illustrate the role of the interacting localized-magnon states 
by comparison with the results considering only independent localized-magnon states, 
i.e., for the lattice-gas models with infinite nearest-neighbor repulsion 
(the one-dimensional hard-dimer and the two-dimensional hard-square model \cite{lm2,independent_lm}). 
The corresponding data are shown in Figs.~\ref{f2} and \ref{f3} by small filled triangles.
The hard-dimer and hard-square model predictions are correct at low temperatures if $h$ is around $h_1$ 
when the independent localized-magnon states do dominate the partition function of the frustrated quantum spin systems.
Clearly, 
including interacting localized-magnon states
a much wider region of validity is obtained in comparison with the hard-core-object description.

\section{Conclusions}
\label{sec4}
\setcounter{equation}{0}

To summarize,
we have considered the low-temperature magnetization processes for two frustrated quantum Heisenberg antiferromagnets.
The possibility to examine these systems in great detail is owing to emergent Ising degrees of freedom in the strong-coupling regime.
As a result, the frustrated quantum systems are similar 
to the classical lattice-gas models or, equivalently, the Ising antiferromagnets in a uniform magnetic field.
Classical lattice systems are much more easier to study 
either analytically (one-dimensional case) or using classical Monte Carlo simulations (two-dimensional case).
The most interesting features of the low-temperature magnetization processes 
are related to the finite-temperature order-disorder phase transition
which occurs in the two-dimensional case.
We have found that the field and temperature dependences of the uniform magnetization and susceptibility 
signalize about ordering of the localized magnons on vertical bonds of the frustrated bilayer. 
The jumps and the plateaus found at zero temperature for both systems 
are typical for strongly frustrated quantum magnets \cite{honecker}.
The steps are smeared out already at quite low temperatures, where the plateau is still visible.
A prominent feature at finite temperatures are the spikes in $\chi(T,h,N)=\partial M(T,h,N)/\partial h$
at the ends of the plateau observed for the ladder system. 
Such spikes have been found theoretically and experimentally, 
for instance, 
for the triangular spin-1/2 Heisenberg antiferromagnet \cite{farnell,ono}
and 
for frustrated magnetic molecules \cite{schroeder}. 
For the bilayer model the sharp peaks in $\chi(T,h,N)$ are related to the phase transition present in this system. 
As a result, they appear for $T<T_0=J_1/[2\ln(\sqrt{2}+1)]\approx 0.567\,296 J_1$, 
i.e., up to quite large temperatures.

Although in the present paper our consideration is restricted to two particular frustrated lattices
it can be obviously extended to other lattices,
for example,
the frustrated three-leg ladder in one dimension \cite{ho_mi_tr,tri}
or frustrated bilayers consisting of two triangular, honeycomb, kagome etc layers in two dimensions.

In general, 
bearing in mind the strong-coupling regime for which our consideration is valid,
we may speak about a system of weakly interacting dimers 
organized in various lattices in different dimensions in such a way 
that the singlet states on the dimers are localized.
Frustrated interactions are necessary to achieve this goal.
From experimental point of view 
(for solid-state realizations of some similar models see Refs. \cite{nedko,nedko_prl})
it would be interesting to consider the case 
when the conditions of localization are slightly violated 
(see, for example, Ref. \cite{mila} 
were magnetization curves for such a frustrated two-leg ladder are discussed at zero temperature).

\section*{Acknowledgments}

O.~D. acknowledges the financial support of the DFG
and thanks Magdeburg University for hospitality in the end of 2010
when the main part of this paper was produced.


\begin{thebibliography}{99}

\bibitem{honecker}
A.~Honecker,
J. Phys.: Condens. Matter {\bf 11}, 4697 (1999);
A.~Honecker, J.~Schulenburg, and J.~Richter, 
J. Phys.: Condens. Matter {\bf 16}, S749 (2004);
J.~Richter, J.~Schulenburg, and A.~Honecker,
in
``Quantum Magnetism'',
U.~Schollw\"{o}ck, J.~Richter, D.~J.~J.~Farnell, R.~F.~Bishop, Eds.
(Lecture Notes in Physics, 645)
(Springer, Berlin, 2004),
pp.85-153.

\bibitem{lm1}
J.~Schnack, H.-J.~Schmidt, J.~Richter, and  J.~Schulenburg,
Eur. Phys. J. B {\bf 24}, 475 (2001);
J.~Schulenburg, A.~Honecker, J.~Schnack, J.~Richter, and H.-J.~Schmidt,
Phys. Rev. Lett. {\bf 88}, 167207 (2002);
for a review see
J.~Richter, J.~Schulenburg, A.~Honecker, J.~Schnack, and H.-J.~Schmidt,
J. Phys.: Condens. Matter {\bf 16}, S779 (2004);
J.~Richter,
Fizika Nizkikh Temperatur (Kharkiv) {\bf 31}, 918 (2005)
[Low Temperature Physics {\bf 31}, 695 (2005)];
O.~Derzhko, J.~Richter, A.~Honecker, and H.-J.~Schmidt,
Fizika Nizkikh Temperatur (Kharkiv) {\bf 33}, 982 (2007)
[Low Temperature Physics {\bf 33}, 745 (2007)];
J.~Richter and O.~Derzhko,
in
``Condensed Matter Physics in the Prime of the 21st Century. Phenomena, Materials, Ideas, Methods'',
J.~J\c{e}drzejewski, Ed.
(World Scientific, Singapore, 2008),
pp.237-270.

\bibitem{lm2}
M.~E.~Zhitomirsky and H.~Tsunetsugu,
Phys. Rev. B {\bf 70}, 100403(R) (2004);
O.~Derzhko and J.~Richter,
Phys. Rev. B {\bf 70}, 104415 (2004);
M.~E.~Zhitomirsky and H.~Tsunetsugu,
Prog. Theor. Phys. Suppl. No. 160, 361 (2005);
O.~Derzhko and J.~Richter,
Eur. Phys. J. B {\bf 52}, 23 (2006).

\bibitem{independent_lm}
J.~Richter, O.~Derzhko, and T.~Krokhmalskii,
Phys. Rev. B {\bf 74}, 144430 (2006);
O.~Derzhko, J.~Richter, and T.~Krokhmalskii,
Acta Physica Polonica A {\bf 113}, 433 (2008).

\bibitem{interacting_lm}
O.~Derzhko, T.~Krokhmalskii, and J.~Richter,
Phys. Rev. B. {\bf 82}, 214412 (2010).

\bibitem{ho_mi_tr}
A.~Honecker, F.~Mila, and M.~Troyer,
Eur. Phys. J. B {\bf 15}, 227 (2000).

\bibitem{bilayer}
P.~Chen, C.-Y.~Lai, and M.-F.~Yang,
Phys. Rev. B {\bf 81}, 020409(R) (2010).

\bibitem{HJS}
H.-J.~Schmidt,
J. Phys. A {\bf 35}, 6545 (2002).

\bibitem{linear}
H.-J.~Schmidt, J.~Richter, and R.~Moessner,
J. Phys. A {\bf 39}, 10673 (2006).

\bibitem{mueller-hartman}
E.~M\"{u}ller-Hartmann and J.~Zittartz,
Z. Phys. B {\bf 27}, 261 (1977).

\bibitem{wu}
X.~N.~Wu and F.~Y.~Wu,
Phys. Lett. A {\bf 144}, 123 (1990).

\bibitem{wang}
X.-Z.~Wang and J.~S.~Kim,
Phys. Rev. Lett. {\bf 78}, 413 (1997).

\bibitem{betts}
S.~J.~Penney, V.~K.~Cumyn, and D.~D.~Betts,
Physica A {\bf 330}, 507 (2003). 

\bibitem{baxter}
R.~J.~Baxter,
Exactly Solved Models in Statistical Mechanics
(Academic Press, London, 1982).

\bibitem{farnell}
D.~J.~J.~Farnell, R.~Zinke, J.~Schulenburg, and J.~Richter,
J. Phys.: Condens. Matter {\bf 21}, 406002 (2009).

\bibitem{ono} 
T.~Ono, H.~Tanaka, H.~Aruga Katori, F.~Ishikawa, H.~Mitamura, and T.~Goto, 
Phys. Rev. B {\bf 67}, 104431 (2003).

\bibitem{schroeder} 
C.~Schr\"oder, H.~Nojiri, J.~Schnack, P.~Hage, M.~Luban, and P.~K\"ogerler, 
Phys. Rev. Lett. {\bf 94}, 017205 (2005).

\bibitem{tri}
M.~Maksymenko, O.~Derzhko, and J.~Richter,
Acta Physica Polonica A {\bf 119}, ??? (2011) (in print)
[preprint ICMP-10-08E].

\bibitem{nedko}
N.~B.~Ivanov,
Condensed Matter Physics (L'viv) {\bf 12}, 435 (2009).

\bibitem{nedko_prl}
G.~Seeber, P.~K\"{o}gerler, B.~M.~Kariuki, and  L.~Cronin,
Chem. Commun., 1580 (2004);
N.~B.~Ivanov, J.~Schnack, R.~Schnalle, J.~Richter, P.~K\"{o}gerler, G.~N.~Newton, L.~Cronin, Y.~Oshima, and H.~Nojiri,
Phys. Rev. Lett. {\bf 105}, 037206 (2010).

\bibitem{mila}
J.-B.~Fouet, F.~Mila, D.~Clarke, H.~Youk, O.~Tchernyshyov, P.~Fendley, and R.~M.~Noack,
Phys. Rev. B {\bf 73}, 214405 (2006).

\end{thebibliography}
\end{document}